\documentclass[12pt]{elsart}
\usepackage{latexsym}
\usepackage{amsfonts,amssymb,amstext}
\renewcommand{\d}{\partial}

\newcommand{\C}{{\ensuremath{\mathbb C}}}

\newcommand{\R}{{\ensuremath{\mathbb R}}}
\newcommand{\CP}{{\ensuremath{\mathbb C}}{\ensuremath{\mathbb P}} }

\newcommand{\vv}{{\ensuremath{\mathbf v}}}
\newcommand{\X}{{\ensuremath{\mathsf X}}}
\newcommand{\Y}{{\ensuremath{\mathsf Y}}}
\newcommand{\y}{{\ensuremath{\mathbf y}}}
\newcommand{\nn}{{\ensuremath{\mathbf n}}}
\newcommand{\PRevD}[3]{{\sl Phys. Rev.} {\bf D#1} (#2) #3}
\newcommand{\PLB}[3]{{\sl Phys. Lett.} {\bf #1B} (#2) #3}
\newcommand{\MPLA}[3]{{\sl Mod. Phys. Lett.} {\bf A#1} (#2) #3}

\newcommand{\IJMP}[3]{{\sl Int. J. Mod. Phys.} {\bf A#1} (#2) #3}
\newcommand{\MA}[3] {{\sl Math. Ann.} {\bf #1} (#2) #3}
\newcommand{\JDG}[3] {{\sl J. Diff. Geom.} {\bf #1} (#2) #3}
\newcommand{\MZ}[3] {{\sl Math.Z.} {\bf #1} (#2) #3}
\begin{document}
\begin{frontmatter}
\title{Three-dimensional strings\break
I. Classical theory}
\author[UAM]{Eduardo Ramos\thanksref{emailed}}

\address[UAM]{Dept. de F\'{\i}sica Te\'orica, C-XI\break
                          Universidad Aut\'onoma de Madrid\break
                         Ciudad Universitaria de Cantoblanco\break
                          28049 Madrid, SPAIN}        

\thanks[emailed]{\tt mailto:ramos@delta.ft.uam.es}

\begin{abstract}
I consider a three-dimensional string theory whose action, besides the
standard area term, contains one of the form $\int_{\Sigma}
\epsilon_{\mu\nu\sigma} X^{\mu} d X^{\nu} \wedge d X^{\sigma}$. In the
case of closed strings this extra term has a simple geometrical
interpretation as the volume enclosed by the surface. The associated
variational problem yields as solutions constant mean curvature
surfaces. One may then show the equivalence of this equation of motion
to that of an $SU(2)$ principal chiral model coupled to gravity. 
It is also possible by means of the Kemmotsu representation theorem,
restricted to constant curvature surfaces, to map the solution space
of the string model into the one of the $\CP^1$ nonlinear sigma
model. I also show how a description of the Gauss map of the surface
in terms of $SU(2)$ spinors allows for yet a different description
of this result by means of a Gross-Neveu spinorial model coupled to
2-D gravity. The standard three-dimensional string equations can also
be recovered by setting the current-current coupling to zero.
\end{abstract}
\end{frontmatter}

\section{Introduction}

The geometry of surfaces have found several application in physics
specially since the advent of string theory as a candidate to describe
QCD in four dimensions or as a ``theory of everything'' \cite{Green}.
Nevertheless, in spite of some spectacular successes from the purely
technical point of view, there is a general consensus that we are
still far away from a phenomenologically realistic theory.  In
particular for the case of QCD, the fact that the Nambu-Goto action
only seems to make sense, due to anomalies, in 26 dimensions induced
Polyakov \cite{Polyakov} to consider an alternative approach based on
the coupling of conformal matter to two-dimensional gravity.  Although
classically both approaches are easily seen to be equivalent for the
case of $D$ bosons coupled to two-dimensional gravity (where $D$
represents the dimension of the target space), Polyakov approach
permitted, through a careful treatment of the Weyl anomaly, to extend
the analysis to the non-critical case. But unfortunately the existence
of the ``infamous'' $c=1$ barrier \cite{ZamI} has not allowed us to
study the physically interesting dimensions, arguably 3 and 4.

The purpose of this paper is to study an alternative three-dimensional
string model with the hope that it will bring new insight into this
difficult subject. The model under consideration, besides the standard
area term contains another one that, for the particular case of closed
surfaces, can be interpreted as the volume enclosed by it.  This
implies, among other things, that this action can be useful to
describe the statistical mechanics of interfaces in three dimensions.
This follows from the fact that the volume term may describe a bulk
contribution whenever the energy density of one of the phases is
different from the other.
 
The plan of the paper is as follows. First I will remind the reader of
some basic notions about the geometry of immersed surfaces in $\R^3$
that will be used in the following.  I will then continue by
introducing the three-dimensional string model action under
consideration. Its equations of motion turn to be the condition of
constant mean curvature for a surface immersed in $\R^3$. I will then
show how it is possible to map this classical problem into the one of
a principal chiral $SU(2)$ model coupled to 2-D gravity. 

In the fourth section I pass to introduce the Gauss map of a surface
in $\R^3$, in order to take advantage of all the machinery already
developed by mathematicians in the subject. In particular it is
possible to map the solution space of our string model into the one of
the nonlinear $\CP^1$ model, by means of the Kemmotsu representation
theorem \cite{Kemmotsu} restricted to surfaces of constant mean
curvature.  Although the relationship between surfaces of constant
mean curvature and the $\CP^1$ nonlinear sigma model has already been
used in the physics literature \cite{ViswanathanI}\cite{Ody}, I
believe that completeness, as well as a slightly different
presentation that will be of later use for my specific purposes,
justify a detailed presentation. I will also comment about the
relationship with affine $SL(2)$ Toda theory \cite{Bakas}, and the
geometrical interpretation of its affine Toda fields
\cite{ViswanathanII}.

In the fifth section I will use the covariant spinorial construction,
developed to introduce the Gauss map \cite{Osserman}, to show the 
equivalence of the
string model with a spinorial Gross-Neveu model. I will then show that
in the limit when the current-current interaction is set to zero one 
recovers the equation of motion of the standard three-dimensional
string action. As a simple exercise, I will recover
the Weierstrass-Enneper representation for minimal surfaces inside
this formalism.

I will finish by making some considerations about the quantization
of this model and remarking some relationships with the Polyakov
rigid string approach \cite{Polyakov}\cite{ViswanathanI}.

\section{A very brief course about surface theory in $\R^3$}

The purpose of this introductory section is to present in a simple
manner the most important geometrical constructions to be used in the
sequel, as well as to set up my notations.  For a comprehensive
introduction to this fascinating subject I refer the reader to the
excellent book of M.~Spivak \cite{Spivak}.

Let $\Sigma$ be an oriented two-dimensional connected Riemannian
manifold and $\X: \Sigma\rightarrow\R^3$ an isometric immersion of
$\Sigma$ into $\R^3$.  At any point $p$ of $\Sigma$ a basis for the
tangent plane is provided by $\partial_{\alpha} X^{i}$. The induced
metric, or first fundamental form of the immersion, is then given by
\begin{equation}
g_{\alpha\beta} = \partial_{\alpha}\X\cdot\partial_{\beta}\X.
\end{equation}
It is now possible to obtain a basis for $T\R^3$ at $p$ by adding a
unitary perpendicular vector $\nn$, which explicit coordinate
expression may be given by
\begin{equation}
n^{i} = {1\over 2\sqrt{g}} \epsilon^{ijk}\epsilon^{\alpha\beta}
\d_{\alpha}X^{j}\d_{\beta}X^{k},
\end{equation}
with $g$ being the determinant of the induced metric.

One may now write down the structural equations of the immersion as
\begin{eqnarray}
\d_{\beta}\d_{\alpha}\X =&&\Gamma^{\rho}_{\beta\alpha}\d_{\rho}\X
+ K_{\beta\alpha}\nn \\
\d_{\alpha}\nn =&& - g^{\beta\rho}K_{\alpha\beta}\d_{\rho}\X. 
\end{eqnarray}

The first of this equation may be taken as the the definition of the
extrinsic curvature $K$, or second fundamental form of the immersion,
while the second follows from consistency with the relations
$\nn\cdot\nn=1$ and $\d_{\alpha}\X\cdot\nn=0$. Notice that multiplying
the first of this equations by $\d_{\gamma}\X$ one readily obtains
that the connection coefficients $\Gamma$ are the ones of the
Levi-Civita connection associated with the induced metric;
multiplication by $\nn$ implies that $K$ is a symmetric tensor.

The Codazzi-Mainardi equation is obtained from
\begin{equation}
\d_{\gamma}\X\cdot (\epsilon^{\alpha\beta}\d_{\alpha}\d_{\beta}\nn )=0,
\end{equation}
which yields that $\nabla_{[\alpha}K_{\beta ]\gamma} =0$.
And finally the Gauss equation is obtained from
\begin{equation}
\d_{\gamma}\X \cdot
(\epsilon^{\rho\beta}\d_{\rho}\d_{\beta}\d_{\alpha}\X)=0,
\end{equation}
which implies that $R_{\gamma\alpha\rho\beta}= K_{\gamma [\rho}
K_{\beta ]\alpha}$, where $R$ is the Riemann curvature tensor
associated with the induced metric.

It is now intuitively clear that given two symmetric tensors $g$ and
$K$ obeying the integrability condition one may recover, up to
Euclidean motions\footnote{This due to the fact that the first and
second fundamental forms, as defined above, are invariant under global
translations and rotations in $\R^3$.}, the associated surface by
integrating the structural equations.

One may define now the mean curvature, $H$, and the Gaussian
curvature, $K$, by
\begin{equation}
H = {1\over 2}g^{\alpha\beta} K_{\alpha\beta},\quad{\rm and}\quad
K = {1\over 2} \epsilon^{\alpha\rho}\epsilon^{\beta\gamma}
K_{\alpha\beta}K_{\rho\gamma}.
\end{equation}

With all of this in mind we now may pass to study the string model at
hand.

\section{The action principle}

As already commented in the introduction I will consider in what
follows a string action that besides the standard area term contains a
contribution of the form
\begin{equation}
S_I = {\Omega\over 3}\int_{\Sigma} d^2 x
\,\epsilon_{ijk}\,\epsilon^{\alpha\beta}
\, X^{i}\partial_{\alpha} X^{j}\partial_{\beta} X^{k}.
\end{equation}

Despite its appearance $S_I$ its invariant under Euclidean motions in
the target space. While the rotational invariance is explicit,
translational invariance is only achieved up to total derivatives,
which of course do not change the dynamical properties of the action.

Does $S_I$ have a simple geometrical interpretation?  The answer turns
out to be positive. This can be most easily seen by rewriting the
extra term in the action as 
\begin{equation}
S_I= {2\Omega\over 3}\int_{\Sigma} d^2 x\,\sqrt{g}\,\, \X\cdot\nn,
\end{equation}
which, in the case of a closed surface, is proportional to the
enclosed volume. This can be easily checked by taking the origin to be
an interior point of the surface and considering the volume element
enclosed by an infinitesimally small solid angle bounded by the
surface. This term has already been considered in the mathematical
literature as a Lagrange multiplier to determine closed minimal
surfaces subject to a constant volume constraint \cite{Barbosa}.

A straightforward computation shows that the equation of motion
associated with the full action $S(\Sigma) = {1/\tau}Area (\Sigma) +
S_I (\Sigma)$ also has a simple geometrical interpretation: the
solution of the associated variational problem is given by surfaces of
constant mean curvature. Explicitly
\begin{equation}
\Box \X = \tau\Omega\, \nn,\label{eq:solution}
\end{equation}
which corresponds to a constant mean curvature $H =\tau\Omega/2$, as
follows from the definition of $H$.  Notice that this result can also
be achieved by choosing the Nambu-Goto or Polyakov prescription
for the area term in the action.

A parenthetical comment: one may choose a representation where
Euclidean invariance is manifest by introducing a vector valued
auxiliary field $\mbox{\boldmath $\varphi$}$.  Let me consider the
following action
\begin{equation}
\tilde S = {1\over\tau}{Area}(\Sigma ) +
{2\over 3}\Omega \int d^2 x \epsilon_{ijk}\epsilon^{\alpha\beta}
\left(\varphi^{i}
\d_{\alpha}X^{j}\d_{\beta}X^{k} -{1\over 2}
\varphi^{i}
\d_{\alpha}\varphi^{j}\d_{\beta}X^{k}\right).
\end{equation}
Now one may eliminate the auxiliary field from the equation of motion
and recover the constant mean curvature condition, while keeping
explicit translational invariance.

Interestingly enough equation (\ref{eq:solution}) can be written as a
zero curvature condition associated with a $SU(2)$ gauge
connection. The construction goes as follows: let me define a one-form
$A$ taking values in the Lie algebra\footnote{I take the adjoint
representation of $SU(2)$ to be spanned by the antihermitian matrices
$i\sigma^j$.} of $SU(2)$.
\begin{equation}
A =i(\tau\Omega )\, \star {\rm d} \X\cdot \mbox{\boldmath $\sigma$},
\label{eq:map}
\end{equation}
where the star stands for the two-dimensional Hodge dual with respect
the induced metric, while the $\sigma_{j}$ are the standard Pauli
matrices. If one works within the Polyakov approach, because the Hodge
star for a one-form is blind to Weyl rescalings of the metric one
could have also used the Polyakov metric.

It follows directly from the definition that $\star\,{\rm d}\star A
=0$, which is tantamount to saying that $A$ is in the Lorenz gauge. It
is a now a straightforward computation to check that the zero
curvature condition
\begin{equation}
{\rm d}A + A\wedge A=0
\end{equation}
reproduces the constant mean curvature condition when written in terms
of $\X$. Therefore, there is a one-to-one relationship between flat
$SU(2)$ connections in the Lorenz gauge and constant curvature surfaces.

One may use now the flatness of $A$ to parametrize it in the usual
way using group variables: $A = h^{-1}{\rm d} h$. But notice
now that the Lorentz condition becomes in terms of $h$ nothing but
the equations of motion of the principal chiral $SU(2)$ model coupled
to 2-D gravity, i.e.
\begin{equation}
\star\, {\rm d} \star h^{-1} {\rm d} h =0,
\end{equation}
which follow from the action
\begin{equation}
{\rm Tr}\int d^2 x \sqrt{g} g^{\alpha\beta} h^{-1}\d_{\alpha} h
h^{-1}\d_{\beta} h.
\end{equation}
A little more of work shows that the equation of motion for the 2-D
metric implies that $g_{ab}$ is conformally equivalent, through the
map (\ref{eq:map}), to the induced one; thus proving the equivalence
between the solution spaces of both theories.

The geometrical data necessary to encode the geometrical properties of
constant curvature surfaces is supplied by the Gauss map of the
surface. Although a thorough introduction to the subject can be found
in a plethora of good textbooks, I will pass now to introduce the
necessary concepts in a way that will show particularly useful for my
purposes.

\section{The generalized Gauss map}

One of the main geometrical tools in the study of immersed surfaces
in $n$-dimensional Euclidean space is provided by the generalized
Gauss map. This map is most simply defined as the map assigning
to any single point in the immersed surface $\Sigma$ its tangent plane, i.e.
it is a map from $\Sigma$ into $G_{(2,n)}$ (the grassmannian of two planes
in $\R^n$). In the case the surface is immersed in $\R^3$
is easy to convince oneself that this map is equivalent to the classical
Gauss map which associates to every point in $\Sigma$ its unit normal vector
\footnote{This requires, of course, the choice of an orientation.}.

There are traditionally several ways to parametrize this map. For our
purposes it will prove convenient to parametrize a tangent plane by a
null complex vector $\vv$ modulo the multiplication by a nonzero
complex number. It is clear that
\begin{equation}
\vv_0 = {\rm Re}(\vv )\quad {\rm and}\quad \vv_1 = {\rm Im}(\vv ),
\end{equation}
form an orthogonal frame with $|\vv_0|=|\vv_1|$ by virtue
of the condition  $\vv\cdot\vv=0$. Notice also that multiplication
of $\vv$ by a complex number simply amounts to a rotation and
dilatation of the frame thus corresponding to the same tangent
plane.

The connection of $G_{(2,3)}$ with $\CP^1$ can be most elegantly made
by using the two-to-one homomorphism between $SU(2)\cong Spin(3)$ and
$SO(3)$. To each vector in $\R^3$ one can associate a $2\times 2$
matrix in the algebra of $SU(2)$ as follows
\begin{equation}
y^i \rightarrow Y^{AB} = y^i\sigma_i^{AB}=
\left(
\matrix{    y^3 &   y^1-iy^2 \cr
            y^1+iy^2  &   -y^3    }\right),
\end{equation}
where the $\sigma_i$ are the standard Pauli matrices; our convention for
them can be easily read from the above formula. From here it follows that
\begin{equation}
|\y|^2= -{\rm det}\,\Y= {1\over 2}{\rm Tr}\, \Y\Y,
\end{equation}
thus making explicit the above homomorphism between Lie algebras, with
the equivalent of the Euclidean metric in $\R^3$ being provided
by $\delta^{AB}$.

From the fact that $\vv = \vv_0 + i\vv_1$ is a null complex vector
is easy to convince oneself that it could be represented in terms
of a complex two-spinor $\xi^A$ by
\begin{equation}
\vv = i\xi^A (\sigma_2\mbox{\boldmath $\sigma$} )_{AB}\xi^B,\label{eq:basic}
\end{equation}
while $\bar\vv$ stands for its complex conjugate. In order
to obtain the above relation we have used in a 
crucial way that
\begin{equation}
\mbox{\boldmath $\sigma$}_{AB}
\mbox{\boldmath $\sigma$}_{CD} =\delta_{AD}\delta_{BC} -
\epsilon_{AC}\epsilon_{BD}.
\end{equation}
From this we have that
\begin{equation}
|\vv_a| = {1\over 2}\vv\bar\vv=(\xi\bar\xi)^2,
\end{equation}
for $a=0,1$.
Moreover, it also directly follows that
\begin{equation}
(\xi\bar\xi)\nn =\bar\xi\mbox{\boldmath $\sigma$}\xi,
\label{eq:normal}
\end{equation}
where $\nn$ is a real unit vector everywhere perpendicular to the 
plane determined by the $\vv$'s.

From the fact that $\vv$ is defined up to multiplication by an
arbitrary nonzero complex number it follows that $\xi^A$ is also
defined modulo multiplication by $a\in\C^{\times}$, thus being
naturally identified with homogeneous coordinates in $\CP^1$. We can
now make contact with standard parametrizations (which otherwise hide
the Euclidean invariance of the target space) by choosing standard
inhomogeneous coordinates. Explicitly by setting $\xi^0=\omega$ and
$\xi^1 =1$ one obtains
\begin{equation}
\nn ={1\over 1+\omega\bar\omega} ( 2 {\rm Re}(\omega), -2 {\rm Im}(\omega),
\omega\bar\omega -1 ).
\end{equation}
The coordinate $\omega$ has otherwise a simple geometrical
interpretation as the complex coordinate associated by stereographic
projection with the Riemann sphere. One should of course be careful
with the fact that these coordinates cannot cover the whole of the
Riemann sphere and should work instead with the standard
atlas.  I will at any rate obviate here the details and leave the
reader to fill the gaps \cite{Kemmotsu}.

The reason for the above exercise is to show how the constant
curvature condition can be neatly written in terms of the Gauss map
and its different parametrizations. Let me start by the classical
Kemmotsu representation of a three-dimensional surfaces in terms of
the mean curvature and the Gauss map.

In order that $\omega (z,\bar z)$ describes the Gauss map of a surface some
integrability conditions must be fulfilled. They are easily stated:
if the $\vv_i$, with $i = 0,1$, are to 
define an orthogonal frame to the surface there must exist a $zweibein$
field such that
\begin{equation}
\partial_{\alpha}\X = e_{\alpha}^i \vv_i.
\end{equation}
Then the integrability condition reduces to
\begin{equation}
0=\epsilon^{\alpha\beta}\partial_{\alpha}\partial_{\beta}\X =
\epsilon^{\alpha\beta}\partial_{\beta}(e_{\alpha}^i \vv_i).
\end{equation}
It will now prove convenient to arrive at a simpler expression via a
judicious use of the symmetries at hand.  The fact that $\vv$ is
defined up to multiplication by a nonzero complex number is translated
into local Weyl and Lorentz invariance of the $zweibein$. One may use
these symmetries together with reparametrization invariance to write
\begin{equation}
\partial\X = e \vv,\label{eq:complexvierbein}
\end{equation}
or equivalently
\begin{equation}
\partial \X = {1\over \sqrt{2}}{e\over 1 + \omega\bar\omega} 
(\omega^2 -1, i(1 + \omega^2), -2\omega),
\end{equation}
where we have fixed the normalization of $e$ such that $\vv$ is complex
unitary, and $\partial$ stands for $\partial /\partial z$. From here it
follows that $\partial \X\cdot \partial\X = \bar\partial\X\cdot
\bar\partial\X =0$, while $g_{z\bar z} = \partial\X\cdot\bar\partial\X
= e\bar e$.

If one now takes the derivative with respect $\bar z$ in 
equation (\ref{eq:complexvierbein}) and projects into the tangent
component one obtains:
\begin{equation}
\bar\partial e + \bar\eta e=0,\label{eq:torsionless}
\end{equation}
where $\eta$ is nothing but the spin connection in the conformal gauge,
i.e. $\bar\eta = \bar\vv \bar\partial \vv$.
The projection into the normal component simply yields an expression
of $e$ in terms of $H$ and $\omega$. Indeed,
from the definition of the mean curvature one obtains that
\begin{equation}
\bar\partial\partial \X \cdot\nn = H e \bar e
\end{equation}
which directly implies that
\begin{equation}
\bar e = {\sqrt{2}\over H}{\bar\partial \omega\over 1 +  \omega\bar\omega},
\end{equation}
where we are assuming that $H$ is everywhere a nonvanishing function. Therefore
one obtains that
\begin{equation}
\partial \X = {\partial\bar\omega\over H}
{1\over (1 + \omega\bar\omega)^2} (\omega^2 -1, i(1 + \omega^2), -2\omega).
\end{equation}
Now the integrability condition of Kemmotsu in terms of $H$ and 
$\omega$ follows from (\ref{eq:torsionless}). A direct computation yields
\begin{equation}
H\left( \bar\partial\partial\omega -2 
{\bar\omega\partial\omega\bar\partial\omega
\over 1 + \omega\bar\omega}\right) =  \partial H\bar\partial\omega.
\label{eq:kic}
\end{equation}
That this is the only integrability condition 
can be checked by a simple counting
argument. In three dimensions one needs three real coordinate functions to
define uniquely a surface, but in the conformal gauge one has 
two extra conditions
coming from $\partial\X\cdot\partial\X =0$, which leaves us with one
degree of freedom, exactly the ones obtained from $H$ and $\omega$ subject 
to the integrability condition (\ref{eq:kic}) above.

All of this allows us to introduce the Kemmotsu representation theorem:

\begin{thm}
Let $\Sigma$ be a simply-connected two-dimensional smooth manifold
and $H: \Sigma\rightarrow \R$ be a nonzero and differentiable function.
Let $\omega : \Sigma \rightarrow S^2$ be a smooth map from the surface into
the Riemann sphere. If $\omega$ satisfies the integrability condition
(\ref{eq:kic}) for the above $H$, then $\omega$ is the Gauss map of
some surface. More precisely, if we define the vector valued
differential form
\begin{equation}
\mbox{\boldmath $\theta$} =  {\rm Re}\left\{ {\partial\bar\omega\over H}
{1\over (1 + \omega\bar\omega)^2} ( \omega^2-1, i(1 + \omega^2), 
-2\omega) dz\right\},
\end{equation}
with $\bar\d\omega\neq 0$ everywhere on $\Sigma$, then
\begin{equation}
\X = \left( \int^z \theta_1 , \int^z \theta_2 , \int^z \theta_3\right)
\end{equation}
describes a regular surface
such that its mean curvature is $H$ and its Gauss map is $\omega$.
\end{thm}

The fact that $\X$ is well defined follows from the closedness of
$\mbox{\boldmath $\theta$}$, i.e.  it is oblivious to the integration path 
used in the
definition,  which is itself a direct consequence of the integrability
condition, as a straightforward computation shows.

After this somehow long detour we may come back to our original
problem. The equations of motion of our string action implied that the
mean curvature was constant. In this case the integrability condition
takes a particularly simple form
\begin{equation}
\bar\partial\partial\omega -2 {\bar\omega\partial\omega\bar\partial\omega
\over 1 + \omega\bar\omega} = 0.
\end{equation}
which is nothing but the equation of motion of the $\CP^1$ nonlinear
sigma model in stereographic coordinates. Now through the Kemmotsu
representation theorem we may obtain a regular surface which is
solution of the string equation by setting $H= \tau\Omega/2$ and
choosing a nonholomorphic solution of the $\CP^1$ model. The reason to
exclude the holomorphic solutions, which correspond to instantons of
the associated nonlinear sigma model, come from the fact that those
surfaces are minimal and therefore have vanishing mean curvature.

It is also of independent interest to notice that by a slight
modification of standard procedures \cite{Bakas} one can show that the
integrability equation takes the form of the affine $SL(2)$ Toda field
equations in the variables
\begin{equation}
{\rm e }^{\phi_1} 
= {\partial\bar\omega\bar\partial\omega \over
(1 + \omega\bar\omega)^2},\quad{\rm and}\quad
{\rm e }^{\phi_2} 
= {\partial\omega\bar\partial\bar\omega \over
(1 + \omega\bar\omega)^2}.
\end{equation}
The Toda variables have now a direct geometrical interpretation
\cite{ViswanathanII}.  The $\phi_1$ field corresponds to the conformal
factor of the induced metric, as can be easily seen by computing
$\partial\X\cdot\bar\partial\X$. One readily obtains
\begin{equation}
g_{z\bar z}= e\bar e = {2\over H^2} {\rm e}^{\phi_1},
\end{equation}
which is up to the constant $2/H^2$ the conformal factor.  The
geometrical interpretation for $\phi_2$ is slightly more involved, but
simple as well. The required geometrical data is provided by the so
called skew curvature; which is nothing but the $K_{zz}$ component of
the extrinsic curvature. If one computes the norm squared of the skew
curvature one obtains
\begin{equation}
K_{zz}K_{\bar z\bar z} = {4\over H^2}
{\partial\bar\omega\bar\partial\omega \over
(1 + \omega\bar\omega)^2}{\partial\omega\bar\partial\bar\omega \over
(1 + \omega\bar\omega)^2}= 2 g_{z\bar z}{\rm e }^{\phi_2},
\end{equation}
from where directly follows an expression of $\phi_2$ exclusively in terms
of the geometrical data associated with the constant curvature
surface.

\section{The covariant spinorial description}

From the previous results it may
seem that the relationship between the string model
and the nonlinear $\CP^1$ sigma model requires a non Euclidean
covariant choice. Therefore it will be interesting to test how far
one can arrive keeping explicitly Euclidean covariance of the target.
As I would like to show now this is indeed possible and the final
result will be provided by a spinorial Gross-Neveu model
coupled to 2-D gravity.

Let me first show how to recover the covariant description of the
$\CP^1$ model from the structural equations of the surface.
If one computes the covariant Laplacian, with respect the induced
metric, acting on $\nn$ one obtains
\begin{equation}
\Box\nn = g^{\alpha\beta}\nabla_{\alpha}\d_{\beta}\nn =
g^{\pi\rho}g^{\beta\alpha}(\nabla_{\beta} K_{\alpha\pi})\d_{\rho}
\X + K_{\alpha\beta}K^{\alpha\beta}\nn.\label{eq:Codn}
\end{equation}
From the Codazzi-Mainardi integrability condition follows that
\begin{equation}
g^{\beta\alpha}\nabla_{\beta} K_{\alpha\pi} -2\nabla_{\pi} H =0,
\end{equation}
thus in the case of constant curvature surfaces one gets that
\begin{equation}
g^{\beta\alpha}\nabla_{\beta} K_{\alpha\pi} =0,
\end{equation}
from where the equation (\ref{eq:Codn}) reduces to
\begin{equation}
\Box\nn = K_{\alpha\beta} K^{\alpha\beta} \nn~.
\end{equation}
A moment's thought reveals that $ K_{\alpha\beta} K^{\alpha\beta}=
\nn\cdot\Box\nn$, and one recovers the standard field equation for the
$O(3)$ nonlinear sigma model, which is well known to be equivalent to
$\CP^1$ through the map (\ref{eq:normal}).  Notice that to deduce this
result one only need to use the fact that $H$ is constant, therefore
it includes the case of minimal surfaces for which $H=0$. It is a
simple exercise to check that consistency in that case with the
structural equations imply the instanton condition \cite{Polyakov}.

Yet a different, and more interesting, formulation is obtained if one
chooses to work in a Euclidean covariant manner within the spinor
formulation.  Let me come back to the covariant expression of the
tangent frame in terms of spinors. The integrability condition may now
be derived from
\begin{equation}
\partial\X= i e \xi \sigma_2\mbox{\boldmath $\sigma$} \xi
\label{eq:Xspin}
\end{equation}
by setting the imaginary part of $\bar\d\d\X$ to zero, i.e.
\begin{equation}
\d\mbox{\boldmath $\bar J$} +  
\bar\d\mbox{\boldmath $J$} =0,
\end{equation}
where $\mbox{\boldmath $J$}= e\xi \sigma_2\mbox{\boldmath
$\sigma$}\xi$.  Therefore the integrability condition can be simply
stated as the conservation of the $\mbox{\boldmath $J$}$ current.

All of this suggest to look for Lagrangian densities where the current
defined above is a conserved quantity. Let me consider the Gross-Neveu
type Lagrangian coupled to 2-D gravity
\begin{equation}
{\mathcal L} = i e \xi\sigma_2\bar\partial\xi + \ c.c. \  +  {1\over 2}
\beta \mbox{\boldmath $J$}\mbox{\boldmath $\bar J$}. 
\end{equation}
Invariance under global Lorentz transformations on the target,
\begin{equation}
\delta_{\rho}\xi = i \mbox{\boldmath $\rho$}\mbox{\boldmath $\sigma$}\xi,
\end{equation}
imply the
conservation of $\mbox{\boldmath $J$}$ as desired.
The associated Euler-Lagrange equation for $\xi$ is given by:
\begin{equation}
\bar\d \xi - {1\over 2}\bar\eta \xi = i\beta \bar e\,\sigma_2  \bar\xi
(\xi\bar\xi ),
\end{equation}
where $\eta$ is the spin connection associated with $e$. The equation of
motion for the {\emph zweibein} yields
\begin{equation}
\bar e = -{i\over 2\beta}{\xi\sigma_2\bar\d \xi\over (\xi\bar\xi )^2}.
\end{equation}
Notice that the first hint that we are on the right track comes from this
equation. If one goes to the gauge where $\xi^0 =\omega$ and $\xi^1 = 1$
the above equation reduces to
\begin{equation}
\bar e = {1\over\beta}{\bar\d\omega\over ( 1 +\omega\bar\omega )^2}
\end{equation}
which turns out to be the expression of the induced {\emph zweibein} in terms
of the Gauss map for constant curvature surfaces if we set
$H =\beta$. Notice that the extra factor of $\sqrt{2}(1 +\omega\bar\omega )$
comes from the different parametrization of $e$; if we use (\ref{eq:Xspin})
to define $g_{z\bar z}$ one gets
\begin{equation}
g_{z\bar z} = 2 e \bar e (\xi\bar\xi )^2,
\end{equation}
and not $e\bar e$ as before. As I will pass to show this turns to be more than a 
coincidence.

From the equation of motion one gets that
\begin{equation}
\bar\d\mbox{\boldmath $ J$} = (\bar\d e + \bar\eta e ) \xi\sigma_2
\mbox{\boldmath $\sigma$}\xi - 2i\beta \bar e e (\xi\bar\xi )
\bar\xi\mbox{\boldmath $\sigma$}\xi.
\end{equation}
The first term is identically zero because of the definition of $\bar\eta$.
If one now plugs back the expression of $\d\X$ in terms of 
$\mbox{\boldmath $J$}$
one recovers the constant curvature condition if one sets
\begin{equation}
\beta = {\tau\Omega\over 2}.
\end{equation}

It is clear from all of this that by setting $\beta$ equal to zero
one recovers the usual string equation. In particular in the
gauge $\xi^0 =\omega$ and $\xi^1 =1$ one obtains that
$\omega $ and $e$ are holomorphic functions. From
this one may recover the Weierstrass-Enneper representation
of minimal surfaces as follows. From equation (\ref{eq:Xspin})
one gets that
\begin{equation}
\X = {\rm Re} \int dz \, e \left( \omega^2 -1, i(1 +\omega^2), -2\omega
\right),
\end{equation}
which is well defined by the holomorphicity of $\omega$ and 
the $zweibein$. If one now implements the conformal 
reparametrization
$z\rightarrow \omega^{-1} (z)$ one may rewrite the equation above as
\begin{equation}
\X = {\rm Re} \int dz \left( z^2 -1, i(1 + z^2), -2 z
\right) \zeta (z),
\end{equation}
with $\zeta (z)= e/\d\omega$ an holomorphic function, thus reproducing
the celebrated Weierstrass-Enneper formula.

\section{Some final comments}

I believe that this trip through the geometry associated with the
string model under study has revealed its intrinsic interest.
Of course, the final test should be provided by the quantum
properties of the string model. Nevertheless, it is to be expected
that its interconnections with classical integrable systems should
pave the way to quantization in the covariant phase space approach.
I hope to come back to this subject in a future publication. 

I would not like to finish without commenting on certain results in
the field of three-dimensional rigid strings which seem to establish
some links with the results obtained here. Although the Polyakov rigid
string does not have as general solution constant mean curvature
surfaces, it was noted by Viswanathan and Parthasarathy
\cite{ViswanathanI} that its action reduces to the one of the $\CP^1$
model for constant curvature surfaces. The equivalence of both
approaches only being complete for the anti-instanton solutions of the
$\CP^1$ model.  How all of this fits into our case is something which
for the time being escapes my understanding, but which I believe is
worth of further study.

\begin{ack}
I would like to thank J.M. Figueroa-O'Farrill for reading a previous
manuscript written in spanglish and suggesting a traslation into
poor english. I would also like to thank J. Roca for many
useful conversations on the subject.
\end{ack}

\end{document}